Title

# Crystal growth, measurement and modeling of the optical activity of α-GeO$_2$ – Comparison with α-SiO$_2$

Alexandra Peña[1,*], Corinne Félix[1], Bertrand Ménaert[1], Benoît Boulanger[1,*]

[1] Univ. Grenoble Alpes, CNRS, Grenoble INP, Institut Néel, 38000 Grenoble, France

alexandra.pena@neel.cnrs.fr

benoit.boulanger@neel.cnrs.fr

Abstract

This work aimed first at growing high quality bulk α-GeO$_2$ crystals in their Quartz iso-structural form (α-SiO$_2$), using a high temperature flux method. By optimizing the flux composition and the geometry and orientation of the seeds, it has been possible to achieve growth yields up to 90 % leading for the first time to bulk single crystals up to 3.5 cm$^3$. Thanks to the optical quality and size of the obtained crystals, the second step of this study was the measurement of the optical activity of α-GeO$_2$ between 0.3 and 2 µm. This gave access to the value of one of the independent components of the gyration tensor ($g_{33}$) as a function of the wavelength. An α-SiO$_2$ slab have been used to validate our methodology. Both crystals show a magnitude of optical activity of 3.4x10$^{-5}$ rad/µm in the near IR range and great variations where the higher values are 3.4x10$^{-3}$ rad/µm and 1.7x10$^{-3}$ rad/µm at 0.3 µm for α-GeO$_2$ and α-SiO$_2$, respectively. The measurements are perfectly described by an empirical model that we propose and which relies on the macroscopic first-order electrical susceptibility and on the pitch of the structural helix compared to the wavelength of light.

*Keywords:*

High temperature solution growth

Chiral inorganic crystals

Optical activity

Gyration coefficient $g_{33}$

## 1. Introduction

The growth of bulk single crystals by hydrothermal and flux methods on the one hand, and of epitaxial layers by chemical vapor deposition, metal-organic chemical vapor deposition or pulsed laser deposition of GeO$_2$ in its rutile and trigonal phases have attracted much attention lately [1-7]. The rutile phase (r-GeO$_2$) is a promising ultra-wide bandgap (UWBG) semiconductor, while the trigonal phase (α-GeO$_2$) is of prime interest for its stable piezoelectric properties at high temperature [8-9]. Note that not all the attempts to grow α-GeO$_2$ led to crystals showing a high thermal stability. The samples obtained by hydrothermal method onto a quartz seed present a phase transition at temperatures lower than 200 °C due to the presence of OH- groups incorporated to the crystals during the growth process [2]. At the opposite, the crystals grown from high temperature solutions by the flux method did not suffer of any phase transition after a long annealing period up to temperatures close to 900 °C [3]. Moreover, the determination of the piezoelectric coefficients $d_{11}$ and $d_{14}$, between 20 and 600 °C, shows only a slightly decrease of the values at high temperatures, which has been attributed to the electric contacts but not to the intrinsic behavior of the α-GeO$_2$ crystals [9].

α-GeO₂ crystal is iso-structural to quartz (α-SiO₂) and belongs to the trigonal class *32* with the chiral space group *P3*$_1$*21* (or *P3*$_2$*21*). Then it belongs to the uniaxial optical class. Since it is an acentric crystal, second-order nonlinear interactions are allowed. The non-linearity of this crystal was predicted by the Density Functional Theory (DFT) and experimentally confirmed by measurements of phase-matched Second-Harmonic Generation (SHG) and Sum- and Difference Frequency Generation (SFG and DFG) [10-11]. The corresponding phase-matching angles were measured with an accuracy of ± 0.5°, which allowed the ordinary ($n_o$) and extraordinary ($n_e$) principal refractive indices to be determined with a precision better than $10^{-4}$ [11]. The helical crystallographic structure of α-GeO₂, as shown in Fig. 1 (top), confers chirality to that crystal, as it is also the case for α-SiO₂, which can lead to Optical Activity [12]. This phenomenon, also called Rotatory Power, leads to the rotation of the direction of the linear polarization of light during the propagation as shown in Fig. 1 (down), and that independently of the birefringence that can exist in the direction that is considered.

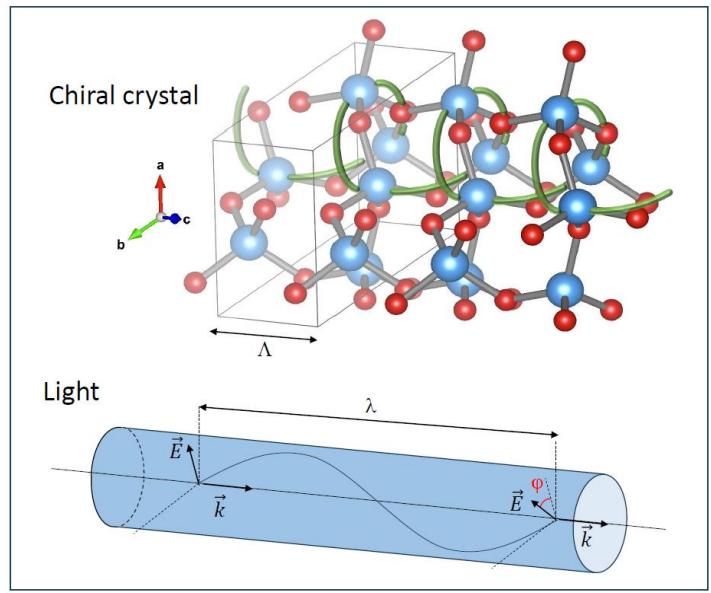

**Fig. 1.** (Top) Chiral crystal structure of α-GeO₂ or α-SiO₂: blue balls represent germanium or silicon atoms, while red balls correspond to oxygens, (**a**, **b**, **c**) is the crystallographic frame and Λ the pitch of the helix. (Down) Scheme of light propagation in a chiral medium: $\vec{E}$ is the electric polarization, $\vec{k}$ the wave vector, $\lambda$ the wavelength, and $\varphi$ the rotation angle of $\vec{E}$ due to optical activity.

In this work, the growth of centimeter size α-GeO₂ crystals using the top-seeded solution growth - slow cooling (TSSG-SC) technique is performed. The good quality of the grown samples allow us to carry out the determination of the magnitude of the optical activity from ultra-violet (UV) to near-infrared (NIR) compared to that of α-SiO₂. We proposed an empirical model that perfectly describes the measurements of both crystals on the basis of the electrical susceptibility, the pitch of the structural helix and the light wavelength.

## 2. Experimental section

*2.1. Chemical reagents*

The growth solutions were prepared by mixing commercial powders of GeO₂ (Fox-Chemicals; 99.999% purity), K₂CO₃ (Fox-Chemicals; 99.9% purity), MoO₃ (Fox-Chemicals; 99.95% purity) and KH₂PO₄ (Merck; 99.5% purity) in the ratio required to obtain solutions of several different molar compositions. The mixed powders, for each of the molar composition studied, were loaded in Platinum crucibles and heated in order to obtain homogeneous solutions from which the α-GeO$_2$ crystals were grown.

*2.2. Crystal growth process*

Top-seeded solution growth by slow cooling (TSSG-SC), which is a high-temperature solution growth method, have been used in this work to grow α-GeO$_2$ single crystals [13,3] in a homemade resistive vertical tubular furnace with three heating zones. The position of the crucible inside the furnace as well as the difference of temperature between the three heating zones are chosen in order to have a small thermal gradient (< 0.5 °C/cm) between the surface and the bottom of the growth solution. To avoid any spurious nucleation during the process, the coldest point of the solution is set at the surface of the solution, *i.e.* at the position where the growth starts onto an oriented seed [14]. After setting the thermal gradient in the solution and before the introduction of the oriented seed, the equilibrium temperature has been found by observing the growth or dissolution of a trial seed. This process is of prime importance in order to precisely set the initial temperature at which the oriented seed will be put in contact with the growth solution. The growth begins when the solution becomes supersaturated and continues while the solution temperature steadily decreases at a rate that goes from 0.02°C/h to 0.20°C/h during the complete growth process that lasts one month. The furnace is equipped with a vertical Z-translation arm, a balance that allows to follow the mass deposition rate, and a rotation engine to fix the suited angular speed of the crystal. The oriented seed is in an off centered position and its rotation is periodically reversed to minimize flow inhomogeneities [15-16]. Once the crystal growth process is over, the crystal is slowly removed from the growth solution and left inside the furnace while it cools down freely to room temperature.

*2.3. Crystal orientation, crystal cutting and technique of measurement of optical activity*

The crystallographic pre-orientation before the cutting of the α-GeO$_2$ seeds, as well as α-GeO$_2$ and α-SiO$_2$ slabs used for the optical measurements, is done by an X-Rays backscattered Laue method using a polychromatic W X-ray source equipped with a 2774x1843 pixels and 14-bit digitization camera from PHOTONIC SCIENCE. After this pre-orientation, with an accuracy better than 0.5°, the cutting is performed with a Diamond wire saw (WELL 3500 Precision): the accuracy of the final orientation is 0.01°, which is achieved by an X-ray SECASI goniometer working in θ-2θ configuration (Cu X-ray source). The optical faces of the slabs used for optical measurements are polished to optical quality with a Logitech PM5 lapping and polishing machine.

The crystalline quality is determined by rocking curve (ω scan) measurement in a SmartLab from Rigaku (Cu X-ray source at 45 kV, 200 mA and K$_{\alpha1}$; step time (st) = 0.15s and step size (ss) = 0.001°). The optical quality of α-GeO$_2$ and α-SiO$_2$, in terms of absence of optical twins is checked between crossed polarizer and analyzer.

The optical activity of the α-GeO$_2$ and α-SiO$_2$ is studied on slabs cut along the z-axis of the dielectric frame, this axis being also the optical axis, i.e. the direction of uniaxial crystals where there is no birefringence. The thicknesses of the α-GeO$_2$ and α-SiO$_2$ slabs are 0.775 mm and 1.785 mm, respectively. The magnitude of the optical activity has been measured with a Lambda 900 spectrophotometer Perkin Elmer controlled by the UV WinLab Software and equipped with automatic rotating polarizers. As shown in Fig. 2, a polarizer and an analyzer have been added in the sample path in order to measure the depolarization of light due to optical activity. This experimental set up was the same than that used for the measurement of the optical activity of YAl$_3$(BO$_3$)$_4$ and K$_2$Al$_2$B$_2$O$_7$ crystals [17].

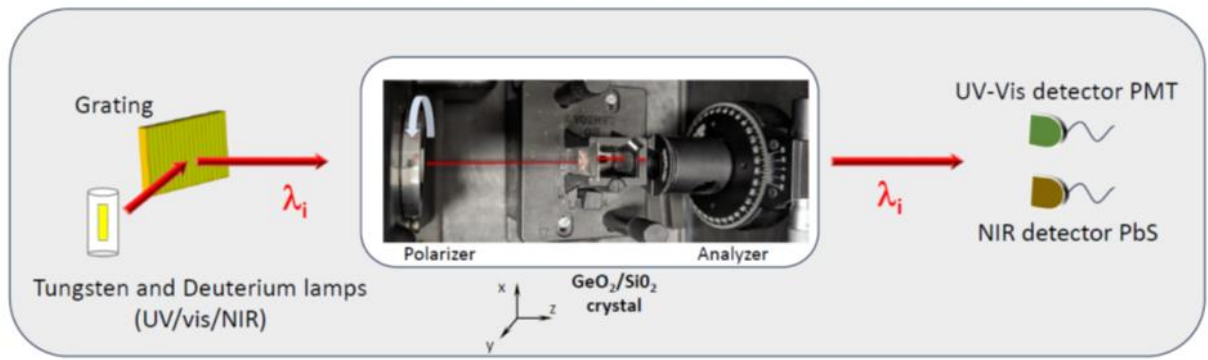

**Fig. 2.** Schematic experimental setup based on a modified Lambda 900 spectrophotometer Perkin Elmer for measuring the magnitude of the optical activity of α-$GeO_2$ and α-$SiO_2$ crystals from 0.3 to 2.0 μm.

## 3. Results and discussion

*3.1. Crystal growth*

The optimization of some experimental conditions, such as the composition of the flux, lead us to the growth of centimeter-size α-$GeO_2$ crystals by the TSSG-SC method without pulling. Commercial $GeO_2$ (c-$GeO_2$) powder, in the hexagonal form, as well as glassy/amorphous $GeO_2$ (g-$GeO_2$) has been taken as solute when using a molybdate solvent; commercial $GeO_2$ has been used in the case of a molybdate-phosphate solvent. With the first solvent, it was not possible to obtain homogeneous solutions with c-$GeO_2$ powder while it is possible with g-$GeO_2$ it is [3]. Even though a homogeneous solution is obtained with the system g-$GeO_2$/molybdate, it suffers from a lack of reproducibility during the crystal growth process. This is due to the relatively high evaporation of the solvent, which leads to a compositional evolution of the growth solution making difficult its implementation. Moreover, the cleaning process of the crucible at the end of the growth is really hard. At the opposite, the molybdate-phosphate solvent allows us to obtain a homogeneous growth solution, a low evaporation of the solvent during the growth, as well as an easy cleaning of the crucible at the end of the process. So, by using a molybdate-phosphate solvent our growth process is more reproducible and easier to implement. In Table 1 are summarized some of the characteristics of the solute-solvent systems described above.

**Table 1**
Comparison of the characteristics of the solute-solvent systems used to grow α-$GeO_2$ crystals

| Solute/solvent system | Solute dissolution | Solvent evaporation | Cleaning of the crucible |
|---|---|---|---|
| c-$GeO_2$/molybdate | partial | Relatively high (> 5%) | Hard (mechanically) |
| g-$GeO_2$/molybdate | complete | Relatively high (> 5%) | Hard (mechanically) |
| c-$GeO_2$/molybdate-phosphate | complete | Relatively low (<2%) | Easy (hot water) |

c-$GeO_2$: commercial hexagonal $GeO_2$ powder
g-$GeO_2$: glassy $GeO_2$

Figure 3 allows us to compare the morphology of two crystals grown with the system g-$GeO_2$/molybdate and the system c-$GeO_2$/molybdate-phosphate. Both crystals were grown by using oriented seed with [001] perpendicular to the surface of the growth solution, and they show {010} and {101} face types. The crystal obtained from the molybdate-phosphate solvent does not show {011} faces. The crystal external morphology changes as a function of the crystal growth parameters, as it is often the case [14, 18]. Here, this change is probably due to the solute concentration or the type of solvent, since the other growth parameters, *e.g.* the axial thermal gradient of the growth solution, the equilibrium temperature and the seed orientation are almost the same. The three principal faces observed in α-$GeO_2$ crystals, *i.e.* {010}, {101} and {011}, are the same that the ones observed in hydrothermally grown α-$SiO_2$ crystals for which the slowest growing faces are {101} ones and the fastest ones are {011} [19-21]. The obtained crystals are quite isometric and the growth rates ranges from 0.25 to 0.45 mm/day along the x-axis, from 0.24 to 0.35 mm/day along the y-axis, and from 0.53 to 0.63 mm/day along the z-axis. The lowest values correspond to crystals obtained with growth yields of about 25% (Fig. 3b) while the growth yields of about 90% lead to the highest values (Fig. 3a). These growth velocities are twice the values obtained previously with the system g-$GeO_2$/molybdate [3].

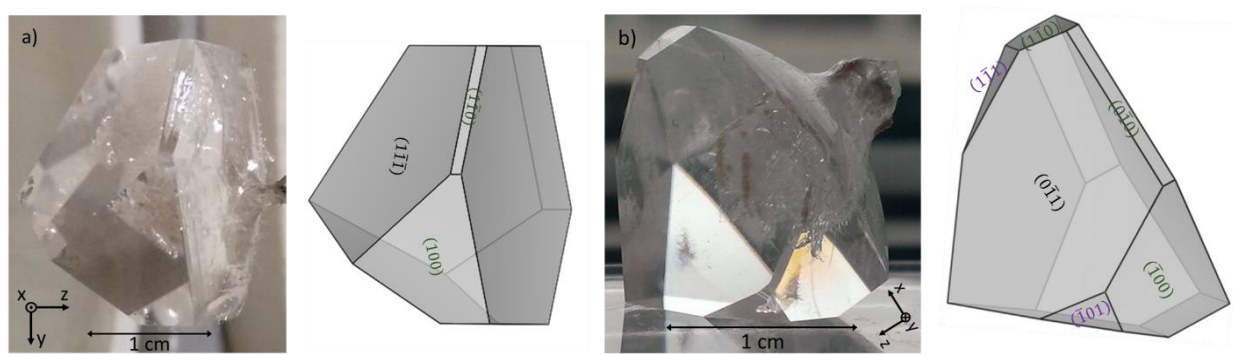

**Fig. 3.** Pictures and corresponding morphologies of two α-$GeO_2$ crystals grown by TSSG-SC by using the system c-$GeO_2$/molybdate-phosphate (a) and the system g-$GeO_2$/molybdate (b). Natural faces present in the as grown crystals have been indexed with Miller indices where black, dark green and violet labels correspond to {101}, {010} and {011} faces respectively [22-23].

*3.2. Crystalline and optical quality*

At naked eye, some macroscopic defects are visible close to the seed position but the rest of the crystal seems to be defect free. A thin α-$GeO_2$ {110} slab have been used for X-ray rocking curve, also called ω-scan, measurements. This method is quite easy to implement and commonly applied to the control of crystalline quality [24-26]. The sharp diffraction peak obtained is centered at 19.548° and have a narrow full width at half-maximum (FWHM) of 0.009° (32 arc sec), attesting of the high quality of the crystals (Fig. 4a).

α-$GeO_2$, as well as α-$SiO_2$, crystallize in the chiral space group $P3_121$ (or $P3_221$). Thus, as α-$SiO_2$ it exists the possibility that α-$GeO_2$ exhibits chiral twins, also called Brazil twins [27]. In crystals having chiral twins, there are zones where the optical axis, which is along the z-axis, can be reversed. These twins can be revealed, characterized and observed by using techniques such as X-ray topography, chemical etching or by using polarized light microscopy [28-30]. In our case, as for the optical activity characterization, only slabs along the optical axis are required, the observation of the possible existence of chiral twins have been done between crossed polarizers. These observations allowed us to select slabs free, or almost free, of twins as the one shown in Fig. 4b.

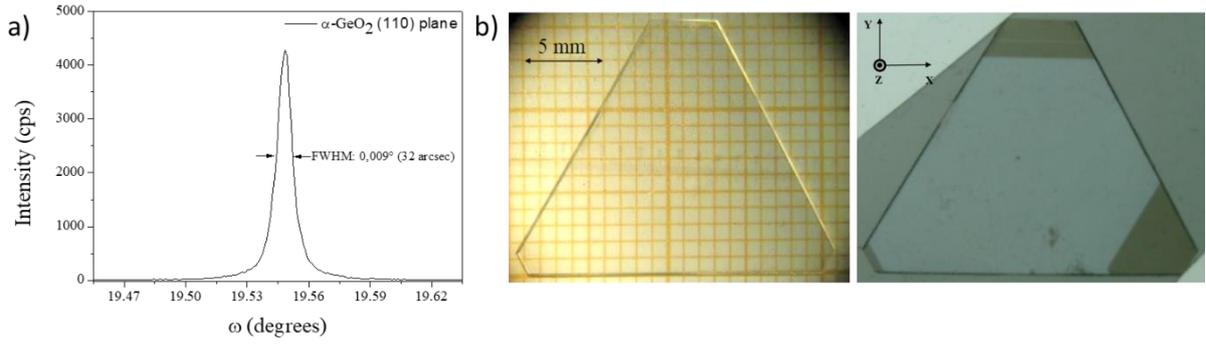

**Fig. 4.** a) Rocking curve on an oriented {110} α-GeO$_2$ slab and b) {001} slab observed under non-polarized light (left) and between crossed polarizers (right). As seen in the right-side image, the centimeter-size {001} slab presents only two twinned zones of few millimeters (dark grey).

*3.3. Measurement and modeling of optical activity*

α-SiO$_2$ and α-GeO$_2$ crystals belong to the uniaxial optical class. Thus, it is along the optical axis of these crystals that optical activity can best be studied, since the birefringence is nil in this direction. The rotatory power ρ(λ) [rad/μm] is then given by the following equation:

$$\rho(\lambda) = \frac{\varphi(\lambda)}{L} = \frac{\pi}{\lambda} \frac{g_{33}(\lambda)}{n_o(\lambda)} \quad (1)$$

where λ [μm] is the light wavelength, L the crystal length, φ the angle of rotation of the polarization of light, n$_o$(λ) the ordinary refractive index, and g$_{33}$(λ) the coefficient of the gyration tensor in the direction that is considered, *i.e.* the optical axis along the crystallographic axis c corresponding to the z-axis of the dielectric frame.

We measured the evolution φ as a function of λ on two samples of α-SiO$_2$ and α-GeO$_2$ with thicknesses L along the crystallographic axis c of 1.785 and 0.775 mm, respectively. The angle of rotation φ is therefore observed in the plane perpendicular to this axis. Measurements were carried out over the wavelength range 0.3 μm ≤ λ ≤ 2 μm. The corresponding values of the rotatory power ρ are plotted in Fig. 5a and compared to existing bibliographic data for α-SiO$_2$ [31-32]. Given these measurements and according to Eq. (1), it was possible to determine the wavelength dispersion of the gyration coefficient g$_{33}$(λ) are given in Fig. 5b knowing the wavelength dispersion of the ordinary refractive index n$_o$(λ) for α-GeO$_2$ and α-SiO$_2$ [11, 33].

The comparison between the spectra of α-SiO$_2$ and α-GeO$_2$ led us to propose a macroscopic model described by the following equation:

$$g_{33}(\lambda) = \Gamma \frac{\Lambda}{\lambda} \left( [n_o(\lambda)]^2 - 1 \right) \quad (2)$$

where Γ is a unitless constant and Λ [μm] the pitch of the structural helix which is none other than the crystal parameter **c**, collinear with the optical axis which is the propagation direction under consideration; the quantity $[n_o(\lambda)]^2 - 1$ is the macroscopic electric susceptibility.

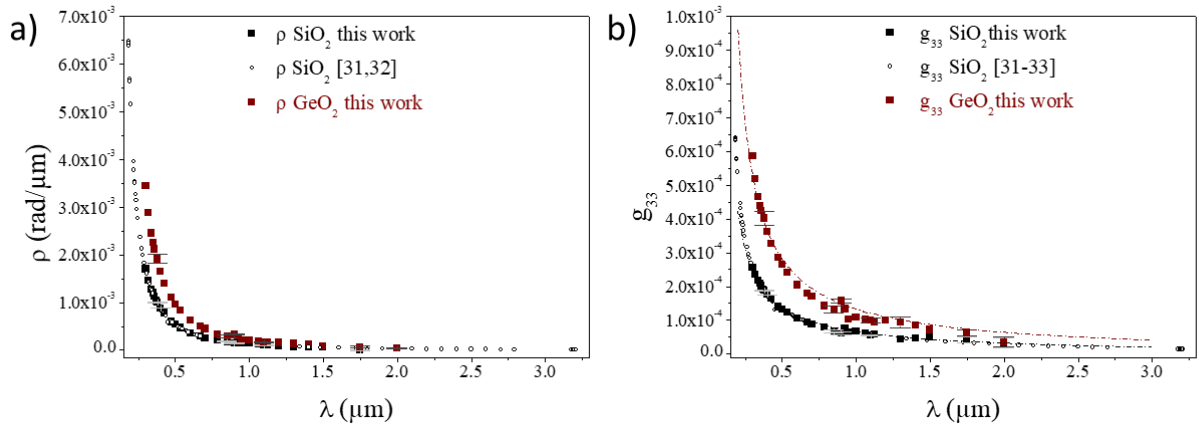

**Fig. 5.** a) Rotatory power ρ measured as a function of wavelength λ for propagation along the optical axis of α-SiO$_2$ and α-GeO$_2$ crystals. b) Wavelength dispersion of the gyration coefficient $g_{33}(\lambda)$.

The interpolation of the experimental data in Fig. 5b by Eq. (2) taking Γ as the interpolation parameter, Λ = 5.4084x10$^{-4}$ µm for α-SiO$_2$ and Λ = 5.6527x10$^{-4}$ µm for α-GeO$_2$, together with the relevant $n_o(\lambda)$ dispersion equations [11, 33] leads to the same value of the Γ constant for both α-SiO$_2$ and α-GeO$_2$, *i.e.* Γ = 0.11 ± 0.02. The fact that the interpolation parameter has the same magnitude for two iso-structural compounds is a good indicator of the validity of our model. In Table 2 are given some particular $g_{33}$ as example. Note the values we measured for α-SiO$_2$ are slightly higher than the ones already published [34]. As it appears in Fig. 5b, $g_{33}$ of α-GeO$_2$ is two times bigger than that of α-SiO$_2$.

**Table 2**

Comparison between several values of g33 values of α-SiO$_2$ and α-GeO$_2$

| Wavelength (µm) | α-SiO$_2$ \|g$_{33}$\| x 10$^{-5}$ | α-GeO$_2$ \|g$_{33}$\| x 10$^{-5}$ | Reference |
|---|---|---|---|
| 0.3 | 24.48 | 52.69 | This work |
| 0.510 | 13.50 | 27.52 | This work |
|  | 12.81±0.08 |  | [34] |
| 0.6328 | 10.74 | 21.64 | This work |
|  | 10.06±0.07 |  | [34] |
| 1.064 | 6.27 | 12.4 | This work |
| 2 | 3.24 | 6.43 | This work |

## 4. Conclusion

α-GeO$_2$ crystals were grown by the top-seeded solution growth – slow cooling (TSSG-SC) method without pulling from two different chemical systems: g-GeO$_2$/molybdate and c-GeO$_2$/molybdate-phosphate. The advantages and drawbacks as well as the morphology of the obtained crystals by using both systems were described. The system c-GeO$_2$/molybdate-phosphate allowed us to grow α-GeO$_2$ bulk single crystals with volumes up to 3.5 cm$^3$ with growth rates up to 0.45, 0.35 and 0.63 mm/day along x-, y- and z-axes, respectively. The X-ray rocking curve measurements attested of the high crystalline quality of the crystals. These results prove that, with the use of the optimized chemical system proposed in this work, large size and good quality α-GeO$_2$ crystals could be obtained and could be used as wave plates or optical filters in conditions of high temperature environments where α-SiO$_2$ crystals could not be used.

A thin oriented z-slab cut in a α-GeO$_2$ crystal free of chiral twins was used to carried out the first measurements of the optical activity of α-GeO$_2$. We proposed an empirical macroscopic model based on a simple geometric approach that describes very well the wavelength dispersion of the phenomenon. This model also applies to the Quartz α-SiO$_2$. The following of this optical study will be to find the physical meaning of the unitless fitting parameter Γ that is equal to 0.11 ± 0.02. For that purpose, the study of α-AlPO$_4$ and α-GaPO$_4$ that are also iso-structural crystals of the quartz family should be of prime interest, which then could lead to establish the generality of our model. The experimental data already published for α-AlPO$_4$ and α-GaPO$_4$ being too scarce to be used [35-36], it will be necessary to characterize these two crystals using the same protocol than that described in the present article.

**CReDiT authorship contribution statement**

**Alexandra Peña:** Conceptualization, Formal analysis, Investigation, Methodology, Writing – original draft. **Corinne Félix:** Investigation, Methodology, Resources. **Bertrand Ménaert:** Funding acquisition, Project administration, Conceptualization, Investigation, Resources. **Benoît Boulanger:** Conceptualization, Formal analysis, Supervision, Writing – original draft, Writing – review & editing.

**Declaration of competing interest**

The authors declare that they have no known competing financial interests or personal relationships that could have appeared to influence the work reported in this paper.

**Data availability**

Data will be available on request.

**Acknowledgements**

Part of this work is in the frame of the OVERHEAT project, which benefited from French government funding managed by the Agence Nationale de la Recherche (21-CE08-0017). The authors are grateful to Jérôme Debray for the preparation of oriented seeds and slabs, and to Stéphane Grenier for the Rocking curve measurements. The LANEF framework (No. ANR-10-LABX-51-01) is acknowledged for its support with mutualized infrastructure.